\documentclass[pre,twocolumn,showkeys]{revtex4} 

\usepackage{graphicx}

\begin{document}

%\title{Quantum Annealing in Kinetically Constrained System}
%\author{Arnab Das, Bikas K. Chakrabarti}
%\affiliation{Saha Institute of Nuclear Physics, 1/AF Bidhannagar,
%Kolkata-700064, India}
%\author{Robin B. Stinhcombe}
%\affiliation{Dept. of Physics, Oxford University, Oxford, England}
%\begin{abstract}
%\end{abstract}
%\maketitle
%\begin{leftline}

\title{Quantum Annealing in a Kinetically Constrained System}

\author{Arnab Das$^1$}
\email[e-mail: ]{arnabdas@cmp.saha.ernet.in}
\author{Bikas K. Chakrabarti$^{1,2}$}
\email[e-mail: ]{bikas@cmp.saha.ernet.in}
\author{Robin B. Stinhcombe$^2$}
\email[e-mail: ]{r.stinchcombe1@physics.ox.ac.uk}

\affiliation{$^1$Saha Institute of Nuclear Physics, 1/AF Bidhannagar,
Kolkata-700064, India}

\affiliation{$^2$Rudolf Peierls Centre for Theoretical Physics,
Oxford University, 1 Keble Road, Oxford, OX1 3NP, UK.}


\begin{abstract}
 Classical and quantum annealing is discussed for 
a kinetically constrained chain of $N$ non-interacting 
asymmetric double wells, represented by Ising
spins in a longitudinal field $h$. It is shown that in certain
cases, where the kinetic constraints may arise from infinitely high but 
vanishingly narrow barriers appearing in the relaxation path of the system, 
quantum annealing exploiting the quantum-mechanical penetration of sufficiently narrow barriers may be far more efficient than its
thermal counterpart. 
 We have used a semiclassical picture
of scattering dynamics to do our simulation for the
quantum system. 
\end{abstract} 

\maketitle

\noindent
\section{Introduction}
Here we demonstrate the effectiveness of quantum annealing [1] in the context
of {\it Kinetically Constrained System} (KCS) [2]. These KCSs are simple 
model systems which have  trivial ground state structures and static
properties, but a complex 
relaxation behaviour due to some explicit constraints 
introduced in the dynamics
of the system [2]. Such systems are very important to understand how much of
the slow and complex relaxation behaviour of a glass can be attributed to its
constrained dynamics alone, leaving aside any complexity of its energy 
landscape structure. 

It has been demonstrated in certain models with energy barriers [1,3] that one can effectively appoint quantum
fluctuations (instead of thermal ones) to anneal a glassy system towards its
ground state. In the method of quantum annealing, one introduces quantum
fluctuations by including a term in the Hamiltonian due to
tunnelling field, that does not
commute with the original (classical) Hamiltonian, and thus generate tunelling
probabilities between the eigenstates (classical configurations) 
of the original (classical) Hamiltonian. 
The introduction of such a quantum tunnelling
is supposed to make the infinitely high but infinitesimally narrow 
barriers transparent to
the system. This allows transitions between different configurations 
classically trapped between
such infinite barriers. In other words, it is expected that 
application of
a quantum tunnelling term will make the free energy landscape ergodic, 
ie the system
will consequently be able to visit any configuration with 
finite probability (Ray et al [1]). 
Finally, of course, the
quantum tunnelling term is to be tuned to zero  to get back 
the ground state of the classical Hamiltonian. 

To study quantum annealing [1,3] in a representative KCS it has to be appropriately generalised.
The KCSs studied so far are all
classical and the constraints are absolutely unsurpassable. To incorporate 
the quantum effect we first visualize that these constraints 
originate from infinitely
high energy barriers, so that the classical system remains unable to jump over such a barrier at any
finite temperature. Then in the quantum version of such a system we
consider the possibility of tunnelling through such barriers quantum 
mechanically
in certain cases when the barrier width approaches zero fast enough
so that the barrier becomes integrable. We specifically define here a quantum 
version of a classical one-dimensional directed KCS, 
known as the East model [4], and study the quantum relaxational 
behaviour and consequent
annealing (to the ground state in the classical limit).

The classical East model is basically a one-dimensional chain of 
non-interacting Ising (`up-down') spins in a longitudinal field
 $h$, say, in downward direction. The ground state of such 
a system is trivially given by all spins down. 
A kinetic constraint is introduced in the model by
putting the restriction that the $i$-th spin cannot flip if the ($i-1$)-th spin
is down. Such a kinetic constraint essentially changes the topology of
the configuration space, since the shortest path between any two 
configurations differing by one or more forbidden flips, is
increased in a complicated manner owing to the blockage of the `straight'
path consisting of direct flips of the dissimilar spins. Further, the constraint becomes 
more limiting as more spins turn down, as happens in the late approach to equilibrium. As a result, the
relaxation processes have to follow more complex and lengthier paths, giving
rise to exponentially large timescale ($\sim e^{1/T^2}$, where $T$ 
is the temperature) [4]. 

\section{Model}
To introduce the possibility of quantum tunnelling through infinite but integrable barriers representing classical
constraints, we start with a chain of
asymmetric double wells, each with a particle localized within it. 
When the barrier (or step) between the two wells is penetrable, then
if we initially prepare a wave packet in one well with sufficiently high
expectation value of kinetic energy $\Gamma$, it will eventually 
tunnel to the other
well. If there is no dissipation, then such to and fro motion between the
 two wells will persist, following successive elastic reflections from the
 infinitely high outer boundaries of the double well. 
 In the corresponding Ising spin representation this tunnelling
dynamics of a particle (wave packet) between the two wells is represented
by a quantum mechanical spin flip dynamics introduced in the model through
the inclusion of a transverse field term in the Hamiltonian.  
Clearly, the value of the transverse field 
$\tilde {\Gamma}$ depends upon the
height and width of the energy barrier
between the wells and the kinetic energy $\Gamma$ of the wave 
packets [5].       
In such a representation, our model reduces to a 
chain of non-interacting Ising spins (double wells) in the presence of a 
downward field $h$ (proportional to the well asymmetry). The spin flip dynamics (flipping probabilities) 
in this model will however be 
calculated directly from a semiclassical picture of the motion of a
particle in a bounded double well.   
The kinetic constraint is introduced by assuming that
the $i$-th spin faces an infinitely high energy barrier between its
two states (up and down), when the ($i-1$)-th spin is down. 
As in the classical model, this barrier is absent when the 
($i-1$)-th spin is up (see Fig. 1). 
When the dynamics is classical, the barriers are impenetrable and the spin at
$i$-th site has to wait until the ($i-1$)-th spin flips to the up state.
In the quantum version
of the model considered here, we allow for tunnelling 
 through such (classically impenetrable)
infinite barriers for the flip of the $i$-th spin even when the ($i-1$)-th spin
is down.  
\begin{figure*}
\resizebox{11.0cm}{!}{\includegraphics{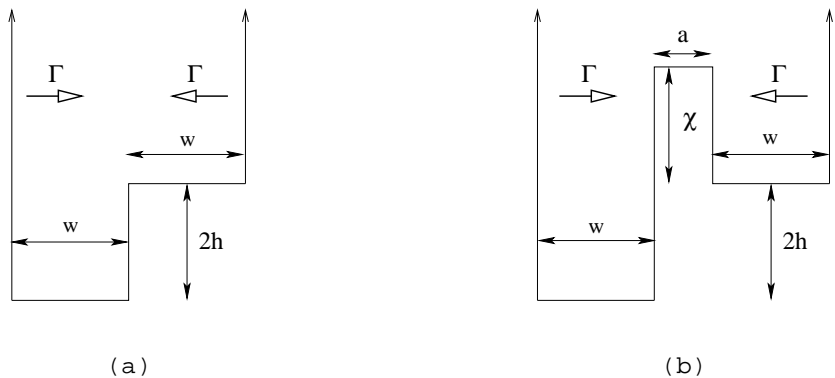}}
\caption{\small{Potential energy wells for the spin at  
site $i$, when ($i-1$)-th
spin is (a) up and (b) down, with the external field $h$ in the downward
direction and barrier height $\chi\rightarrow\infty.$ For the classical 
East model, flipping to right well in (b) is impossible (at any finite $T$).
In the quantum model considered here, although $\chi\rightarrow\infty,$ the
well width $a\rightarrow 0$ in such a way that
 $\chi^{2}a$ $(\equiv g)$ is finite,
giving a finite tunnelling probability for going to the right well in (b).
 The limit w $\rightarrow \infty$ helps utilising
 the scattering picture employed here.}}
\end{figure*}
%To introduce this quantum tunnelling into the
%model, we assume that the kinetic energy  
%$\Gamma$ of a particle in a potential well, 
%which does not commute with the potential energy of
%the particle, 
% allows it to go uphill (i.e., the localized particle or the spin can flip up) 
%by scattering over a potential step, 
%or penetrating through a potential barrier. We also suppose that though in
%the constrained case the classically impenetrable barrier has its height
%$\chi\rightarrow \infty$, it is still integrable, i.e., its width
%$a\rightarrow 0$ in such a manner that $g \equiv \chi^{2}a$
%is finite. In such a case the barrier becomes 
%quantum mechanically penetrable.
The tunnelling
probabilities come from the following semiclassical picture of
scattering of a particle in a double well with
infinitely remote outer boundaries (w $\rightarrow \infty$ in Fig. 1).
If a particle is put in one of the wells of such a double well with some
kinetic energy (actually the expectation value) $\Gamma$, then it will
eventually be scattered by the 
separator (a barrier or step) between the two wells. 
In such a scattering,
there is a finite probability $P$ that the particle manages 
to go to the other
well. We calculate $P$ from the simple picture of scatterings of 
a particle by one dimensional potentials as prescribed below. 
The minimum of the energy of the Ising chain (eqivalent to 
the potential energy of the chain of the double wells) trivially
corresponds to the state
with all the spins down, i.e., aligned along the longitudinal field $h$ 
(where all the particles are in their respective lower wells). 
However, if one starts with
a random configuration and kinetic energy $\Gamma$ is not sufficient for
tunnelling to the upper well, then the system,
more or less, will 
exhibit the zero temperature (energy minimization) relaxation behaviour
of the classical East model, and will extremely slowly approach the ground state
(i.e., the minimum of the potential energy).  
For sufficiently high $\Gamma$, the system 
occasionally tunnels through the infinite barriers corresponding to the
constraints and 
thus can take up some of the  relaxation paths forbidden classically.
However, at any nonzero $\Gamma$, the 
 ground state (lowest potential energy state) will be mixed with higher 
potential energy eigenstates. To reach the ground state, we start with a 
very large initial value of $\Gamma$ and 
then reduce it following an exponential schedule given by 
$ \Gamma = \Gamma_0\exp{(-t/\tau_Q)}.$ Here $t$ denotes the time, and 
$\tau_Q$ sets the effective time scale of annealing. 
At zero temperature the slow spin flip dynamics
occurs only due to the tunnelling (kinetic energy) term
$\Gamma$, and hence the system ceases to have any relaxational dynamics in the 
limit $\Gamma\rightarrow 0 $. It may be mentioned here that
 in absence of any analytical expression for the tunnelling probability
in asymmetric case of the type discussed here, (see e.g., [6]), we employ
the asymmetric barrier tunnelling probabilities available [7].   

\section{Simulation and Results}   
We have employed the quantum transmission (flipping) probablities
(cf. [7])  from a very 
elementary scattering picture which is qualitatively adequate, though not strictly valid for 
the asymmetric double well (shown in Fig. 1(b)) because of its bound states and finite w.
Following are the flipping probabilities 
 ($P$) for the $i$-th spin in different possible situations
used in our Monte Carlo simulation: \\
I.\quad If the $(i-1)$-th spin is up and the $i$-th spin is also up, then
$P = 1$. \\
II.\quad If the $(i-1)$-th spin is up and the $i$-th spin is down, then
(a) $P = 0  $ for $\Gamma < 2h$, and 
(b)$P =min \{1, 4{[\Gamma(\Gamma-2h)]}^{1/2}/(\sqrt{\Gamma} + 
\sqrt{\Gamma-2h})^2\}$, for $\Gamma \ge 2h$. \\
III.\quad  If the ($i-1$)-th spin is down and the $i$-th spin is up then
$P = min\{1, 4{[\Gamma(\Gamma+2h)]}^{1/2}$ 
$/((\sqrt{\Gamma} + \sqrt{\Gamma+2h})^2 + g^2) \}.$\\  
IV.\quad If the ($i-1$)-th spin is down and the $i$-th spin is up then
(a) $P = 0  $ for $\Gamma < 2h$, and 
(b)$P = min\{1, 4{[\Gamma(\Gamma-2h)]}^{1/2}/((\sqrt{\Gamma} 
 + \sqrt{\Gamma-2h})^2 + g^2)\}$ for
$\Gamma \ge 2h $ ($h$ and $\Gamma$ denoting the magnitudes only).\\
 Here $g = \chi^{2}a$, $\chi$ and $a$ being respectively the height 
and width of the barrier representing the kinetic constraint.
The above expressions for $P$ are actually the transmission
coefficients in respective cases of one-dimensional scattering across 
asymmetric barrier or step (according to the form of the potential encountered
in passing from one well to the other, see e.g., [7]).  
Application of the above scattering picture, 
even for the double wells in Fig. 1b
(which our simulation is based on) as discussed before, is of course an 
approximation.

In our simulation, we take $N$ Ising spins
($\sigma_{i} = \pm 1,\quad i = 1,..., N $) on a linear chain with
periodic boundary
condition. The initial spin configuration is taken to be random such that
magnetization $m = (1/N)\sum_{i}\sigma_{i}$ is practically negligible
($m_i \approx 0$).
 We then start
with a tunnelling field $\Gamma_0$ and follow the 
zero temperature (semi-classical)
Monte Carlo scheme as mentioned above, using the spin flip probabilities $P$'s
appropriate for the four cases I-IV. 
Each complete run over the entire lattice
is taken as one time unit and as time progresses, $\Gamma$ is decreased from
its initial value $\Gamma_{0}$ according to 
$\Gamma = \Gamma_{0}e^{-t/\tau_{Q}}$. The results are shown in Fig. 2.
It shows that for $g = 100$ and $\Gamma_{0} = 100$
 the system freezes before reaching the 
ground state ($m_f = 1$) for  low values of $\tau_Q$; say for $\tau_Q = 2000$. 
For a somewhat greater value, e.g., $\tau_Q = 5000$, the system is completely
annealed to the ground state within about $4\times 10^4$ time steps. However,
for a much greater $\tau_Q$, like $\tau_Q = 20000$, the system 
of course anneals completely but consumes more time unnecessarily. 
These generic features remain
the same for other higher values of $g$. We have also studied the
dependence of annealing behaviour with the parameter $g$, which is actually
a measure of how impenetrable is the infinite barrier representing the
kinetic constraint. Computations were carried out 
to locate, for a given value of $g$, the minimum value of $\tau_Q$ for which the
system just anneals upto $m_f = 0.8$ 
(complete annealing requires prohibitively longer computer time for this
comparative study). 
We call this minimum value $(\tau_Q)_{min}$.
A bisection scheme was used to locate $(\tau_Q)_{min}$ for different
values of $g$ starting for the same initial configuration. 
The inset in Fig. 2 shows that
$(\tau_Q)_{min}$ increases fairly sharply with $g$ 
(an empirical analysis shows $(\tau_Q)_{min} \sim ag^2 + b$,
 where $a$ and $b$ are
constants) for $g \le 1000$. 
This quadratic variation with $g$ might be due to the 
specific functional form of $P$ we have used. 
However, for even higher values of $g$, the slope is expected to
 decrease, and finally in the
asymptotic limit $g\rightarrow\infty$, the relaxation behaviour should converge
to that of one with an unsurpassable
kinetic constraint (like the classical East model).
 This asymptotic convergence could not however be explored, 
since the required computational time becomes prohibitively 
long as $g$ is increased further.   

\begin {figure*}
\resizebox{10.0cm}{!}{\includegraphics{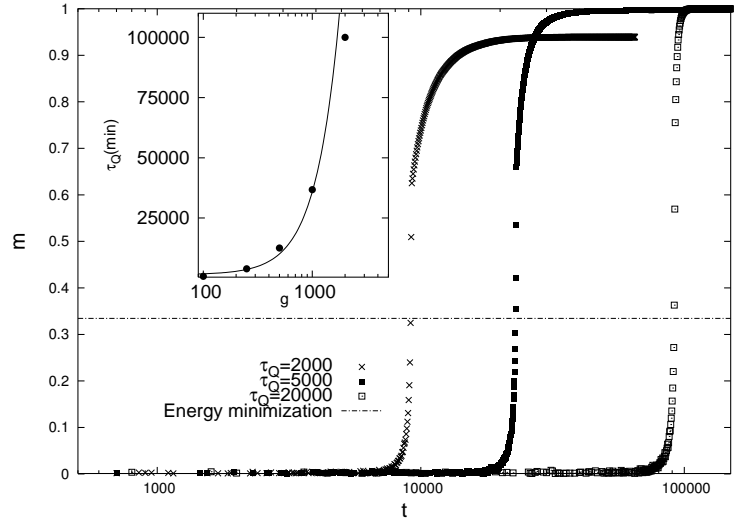}}
\caption{\small{Quantum annealing ($T=0$) for $g=100$,
$\Gamma_0=100$ and $h=1$ is shown for different 
values of $\tau_Q$, for a chain of 5$\times 10^{4} $ spins 
($m$ averaged over the same set 
${\mathcal C}$ of 10 initial configurations for each $\tau_Q$).
The horizontal (dashed) line indicates the average (over the same set
${\mathcal C}$) value of $m$ that could be reached from the initial 
configurations by simply minimizing the energy 
following the downhill principle.
In the inset, variation of $(\tau_Q)_{min}$ with $g$ is shown 
(by the points) for one given
configuration. The error in $(\tau_Q)_{min}$ 
is typically less than $0.5$
\% in each case. The continuous line in the inset shows a fit 
 $(\tau_Q)_{min} = 0.34364g^2 + 1500$.}}
\end{figure*}

\begin{figure*}
%\resizebox{12.0cm}{!}{\rotatebox{270}{\includegraphics{EastC_10L.eps}}}
\resizebox{10.0cm}{!}{\includegraphics{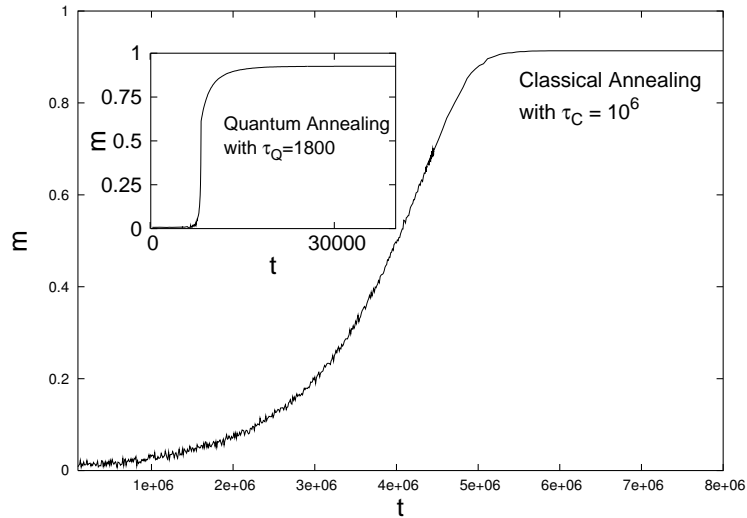}}
\caption{\small{Comparison between classical and quantum
annealing for a chain of $5\times 10^4$ spins (for the same initial disordered
configuration with $m_i \sim 10^{-3}$). We show the results for
 $\tau_Q = 1.8\times 10^{2}$ (for quantum) and 
$\tau_C = 10^{6}$ (for classical) with $h=1$; a lower
$\tau_C$ would not produce substantial annealing.  
Starting from the same initial values $\Gamma_{0} = T_{0} = 100$, 
(and $g=100$
 in the quantum case) we observe that classical annealing requires about
$10^{7}$ steps, whereas quantum annealing takes about $10^{4}$ steps
for achieving the same final order $m_f\sim 0.92$.}}
\end{figure*}

We have also studied thermal annealing of the same (classical East) model
for $\Gamma = 0,$
following an exponentially decreasing temperature schedule given by
$ T = T_0\exp{(-t/{\tau_C})}$; 
 $\tau_C$ being the time constant for the
thermal annealing schedule and $T_0$ the initial temperature.
 Here, when $(i-1)$-th spin is down, the flipping probability for the $i$-th
spin ($\sim \exp{-(\chi/T)}$) is negligible since $\chi$ is very large.
 Otherwise, it flips with probability $P = 1$ if it were in the up state, and 
 with Boltzmann probability $P = \exp{(-h/T)}$ if it were in the down state.
 In Fig. 3 
we compare the results for the same order of initial value and time constant
 for $\Gamma$ and $T$ (barrier height $\chi$ is 
taken to be $1000$ in both the cases while
$g$ was taken to be $100$ in the quantum annealing case,
 or equivalently the barrier width $a$ is taken to be
of the order of $10^{-4}$).
 We observe that to achieve a similar degree of
annealing (attaining a certain final magnetization $m_f$),
 starting from the same disordered configuration, one typically requires much
smaller $\tau_Q$ compared to $\tau_C$; typically, 
$\tau_C \sim 10^{3}\times\tau_Q $ for equivalent annealing 
(for similar optimal values of final order $m_f \sim 0.92$). For annealing
with final order $m_f \sim 1$, we find $\tau_C \sim 10^{4}\times\tau_Q$. This
comparison of course depends on the value of $g$ used (for the barriers)
as shown in the inset of Fig. 2.

\section{Summary}  
We have discussed here the annealing of a kinetically constrained Ising
spin chain of $N$ spins, starting from a disordered 
state (with negligible initial magnetization),
to its (external field induced) fully ordered ground state.
At any finite temperature $T$ (in the classical model) the system takes an
exponentially long time to relax to the ordered state because of the kinetic
constraints, which act like an infinite potential barrier, depending on the
neighbouring spin configurations. 
Quantum mechanically, this infinite
barrier is taken to be penetrable, ie with finite tunnelling probability, depending
on the barrier height $\chi$ and width $a$ ($a \rightarrow 0$ faster than $\chi^{-2}$).
 The introduced noise, required for the annealing,
 is reduced following an exponential schedule in both the cases: 
 $T = T_{0}e^{-t/\tau_C}$, $ \Gamma = \Gamma_{0}e^{-t/\tau_Q},$ with 
$T_0 \approx \Gamma_0$. For our simulation for the quantum case,
 we have taken the tunnelling 
probabilities $P$ (for cases I-IV) and employed them in a 
semi-classical fashion
for the one dimensional spin chain considered.
We observe that for similar achievement
in final order
($m_f \simeq 0.92$ starting from $m_i = 10^{-3}$), 
$\tau_C \sim 10^{3}\tau_Q$ for
$N = 5\times 10^{4}$.  For even larger order ($m_f\sim 1$), quantum annealing
works even better ($\tau_C \sim 10^{3}\tau_Q$, for the same value of $N$).
These comparison  are for $g = 10^2$ and $\chi = 10^3$ for the constraint
barriers. 

In this picture, we considered the collective dynamics of a many particle
system, where each one is confined in a (field) induced asymmetric double
well potential for which we considered only the low lying 
 two states (the wave packet localized in
one well or the other), representing the two states (up and down)
of an Ising spin discussed above.
The tunnelling of the wave packet from one well to the other
was taken into account  by employing a scattering picture and we used the
tunnelling probabilities as the flip probabilities
for the quantum Ising spins.
As such, the reported simulation for the one dimensional quantum
East model is a semiclassical one. It may be noted however that, 
because of the absence of inter-spin interaction, the dimensionality
actually plays no role in this model except for the fact that the
 kinetic constraints on any spin depend only on the left nearest
neighbour (directedness in one dimension).
 Hence the semiclassical one dimensional simulation, instead of
a proper quantum Monte Carlo simulation
 (equivalent to a higher dimensional classical one
 [5]), is quite appropriate here.  
         
\begin{acknowledgments}
B.K.C. is grateful to 
INSA - Royal 
Society Exchange of Scientists Programme for his visit to 
Oxford University, where the work was initiated. 
R.B.S. acknowledges EPSRC support under the grants GR/R83712 and GR/M04426.
\end{acknowledgments}

\bigskip

\noindent {\Large{\bf References}}\\ 

[1] T. Kadowaki and H. Nishimori, Phys. Rev. E {\bf 58} 5355 (1998); 
J. Brooke, D. Bitko, T. F. Rosenbaum, and G. Aeppli, 
Science {\bf 284} 779 (1999); P. Ray, B. K. Chakrabarti 
and A. Chakrabarti, Phys. Rev. B {\bf 39} 11828 (1989) 

[2] G. H. Fredrickson and H. C. Andersen, 
Phys. Rev. Lett. {\bf 53} 1224 (1984); G. H. Fredrickson and H. C. Andersen,
J. Chem. Phys. {\bf 83} 5822 (1985)

[3] G. E. Santoro, R. Martoák, E. Tosatti and R. Car, 
Science {\bf 295} 2427 (2002); R. Martonák, G. E. Santoro and E. Tosatti,
Phys. Rev. E {\bf 70} 057701 (2004)

[4] J. Jackle and S. Eisinger, Z. Phys. B {\bf 84} 115 (1991);
M. A. Munoz, A. Gabrielli, H. Inaoka and L. Peitronero, Phys. Rev. E {\bf 57}
 4354 (1998); F. Ritort and P. Sollich, Adv. Phys., {\bf 52} 219 (2003)

[5] B. K. Chakrabarti, A. Dutta and P. Sen,
{\it Quantum Ising Phases and Transitions in Transverse Ising Models},
 Springer-Verlag, Heidelberg (1996), P. G. de Gennes Solid State Comm. 
{\bf 1} 132 (1963) 

[6] J. G. Cordes and A. K. Das, Superlatt. Microstr., {\bf 29} 121
 (2001), R. Koc, D. Haydargil, arXiv:quant-ph/0410067 v1, (2004)

[7]  H. Margenau and G. M. Murphy,
{\it Mathematics of Physics and Chemistry} (2nd Ed.), 
Van Nostrand, New Jersey (1956), pp 356-358; 
 E. Merzbacher,
{\it Quantum Mecahnics} (3rd Ed.), John Wiley \& Sons Inc.,
USA (1998), pp 93-96

%[8] M. J. de Oliveira and J. R. N. Chiappin, Physica  A {\bf 238} 307 (1997)

\end{document}